\documentclass[%
 aip,
 amsmath,amssymb,
 reprint,%
]{revtex4-2}
\usepackage{epsfig,graphics,amssymb,amsmath,subeqnarray,color,bm,bbm}
\usepackage[toc,page]{appendix}
\usepackage{mathtools}
\usepackage{xcolor}
\usepackage{float}
\usepackage{graphicx}
\usepackage[colorlinks=true,allcolors=blue]{hyperref}

\usepackage{todonotes}

\newcommand{\aver}[1]{\left\langle {#1}\right\rangle}
\newcommand{\beq}{\begin{equation}}
\newcommand{\eeq}{\end{equation}}
\newcommand{\beqa}{\begin{eqnarray}}
\newcommand{\eeqa}{\end{eqnarray}}

\newcommand{\Rar}{{\Rightarrow}}
\newcommand{\dt}[1]{{{\rm d} #1 \over {\rm d} t }}
\newcommand{\cP}{{\cal P}}
\newcommand{\cR}{{\cal R}}

\begin{document}
\title{Stochastic effects on the dynamics of an epidemic due to population subdivision}
\author{Philip Bittihn}
\email{philip.bittihn@ds.mpg.de}
\affiliation{Max Planck Institute for Dynamics and Self-Organization (MPIDS), 37077 G\"ottingen, Germany}
\author{Ramin Golestanian}
\email{ramin.golestanian@ds.mpg.de}
\affiliation{Max Planck Institute for Dynamics and Self-Organization (MPIDS), 37077 G\"ottingen, Germany}
\affiliation{Rudolf Peierls Centre for Theoretical Physics, University of Oxford, Oxford OX1 3PU, United Kingdom}
\date{\today}

\begin{abstract}
    Using a stochastic Susceptible-Infected-Removed (SIR) meta-population model of disease transmission, we present analytical calculations and numerical simulations dissecting the interplay between stochasticity and the division of a population into mutually independent sub-populations. We show that subdivision activates two stochastic effects---extinction and desynchronization---diminishing the overall impact of the outbreak, even when the total population has already left the stochastic regime and the basic reproduction number is not altered by the subdivision. Both effects are quantitatively captured by our theoretical estimates, allowing us to determine their individual contributions to the observed reduction of the peak of the epidemic.
\end{abstract}

\maketitle

\begin{quotation}
    Simple models for the spread of infectious diseases are useful for the quantitative characterization of an epidemic as well as for forecasting future infection numbers and guiding decision-making for containment. Different extensions and refined versions of these models have been created to extract various factors that may be critical for the dynamics and the prevention of epidemics. Although it is well-known that stochastic fluctuations can alter the dynamics as well, they are often neglected at higher infection number levels, such that the contact rates and basic reproduction number become the central quantities of interest. In contrast, we investigate a situation in which stochastic effects can quantitatively change the course of an epidemic when infection numbers are large and contact rates remain unaltered. We consider an extended SIR model in which a large population is subdivided into a certain number of sub-populations, each containing only a few infected individuals. For the limiting case of perfect isolation, i.e. when the epidemic evolves independently in each sub-population with no cross-infections, we derive analytical estimates for these stochastic effects that together recapitulate the results of extensive numerical simulations. Our central quantity of interest is the peak total number of simultaneously infected individuals, which we compare between the subdivided population and a single large population with identical reproduction number. Our analysis suggests that regional isolation can resurrect certain stochastic effects and thereby contribute to effective containment, regardless of the initial distribution of infected individuals.
\end{quotation}

\section{Introduction}

Generic models such as the Susceptible-Infected-Removed (SIR) model conceived by Kermack and McKendrik \cite{Kermack:1927cy} are indispensable for characterizing the bulk properties of epidemics and determining the influence of crucial parameters on the dynamics. The contact rate between individuals, which is proportional to the reproduction number $\cR_0$, usually plays a crucial role, as its reduction through containment measures directly slows the spreading of the disease. On a large scale (states or countries), numbers of infections during the height of an epidemic are usually large, such that deterministic mean-field descriptions are appropriate. These have been widely used to track the course of epidemics and the effect of interventions, for example for the current spreading of COVID-19~\cite{Dehning:2020gk}.

While many details about the biology and modes of infection of a specific disease are important for its dynamics in detailed models~\cite{Heesterbeek:2015de}, even basic SIR models have been extended in various conceptual directions. Besides various general topologies of the underlying contact and mobility networks \cite{Hufnagel:2004kt,Matamalas:2018ck,PastorSatorras:2015gk}, so-called meta-population models have been used to separate the disease dynamics \emph{within} local environments from its spread \emph{between} them~\cite{Gilpin:1991ib}. It has been shown that it is possible to calculate effective quantities for the whole population, such as reproduction numbers (i.e. a threshold theorem)~\cite{Dushoff:1995ht}, the final attack ratio~\cite{Lunelli:2018ky} and criteria for persistence~\cite{Andreasen:1989hf} in deterministic models of such subdivided populations.

Another important important deviation from simple bulk behavior arises through stochasticity (see Ref.~\citenum{Andersson:2000wk} and references therein). Stochastic versions of extended SIR and related models have been used to calculate corrections to the outbreak threshold~\cite{Hartfield:2013kx}, consequences of stochasticity for contact tracing~\cite{Huerta:2002fe} and other control schemes~\cite{Allen:2017jm}, to only name a few. Stochastic effects are also observed in agent-based~\cite{Tiwari:2020cf} and meta-population models~\cite{Keeling:2000gf,Grenfell:1997gt}. Here we seek to study the joint effect of subdivision and stochasticity on the overall magnitude of an epidemic for a fixed initial number of infected individuals in the total population.
In general, subdivision can be expected to artificially boost fluctuations, as the infection numbers in each sub-population can be small even when the total number of infections in the entire population is large. We would like to quantify the ability of such increased stochasticity to reduce the impact of the epidemic. We deliberately refrain from applying any form of traditional containment in our model, such as further reductions in contact rate or contact tracing\footnote{The combination of traditional containment measures with relative regional isolation is the subject of a separate study~\cite{Bittihn:2020do}}. In particular, we design the subdivision such that the deterministic dynamics of the epidemic in the subdivided population remains unchanged compared to a single large population, as outlined in section \ref{sec:model}. This allows us to compare the peak number of infected individuals in the entire population for each scenario both analytically and numerically in order to extract the specific effects of stochasticity triggered by  subdivision.

\begin{figure}[t]
\centering
\includegraphics[width=0.93\columnwidth]{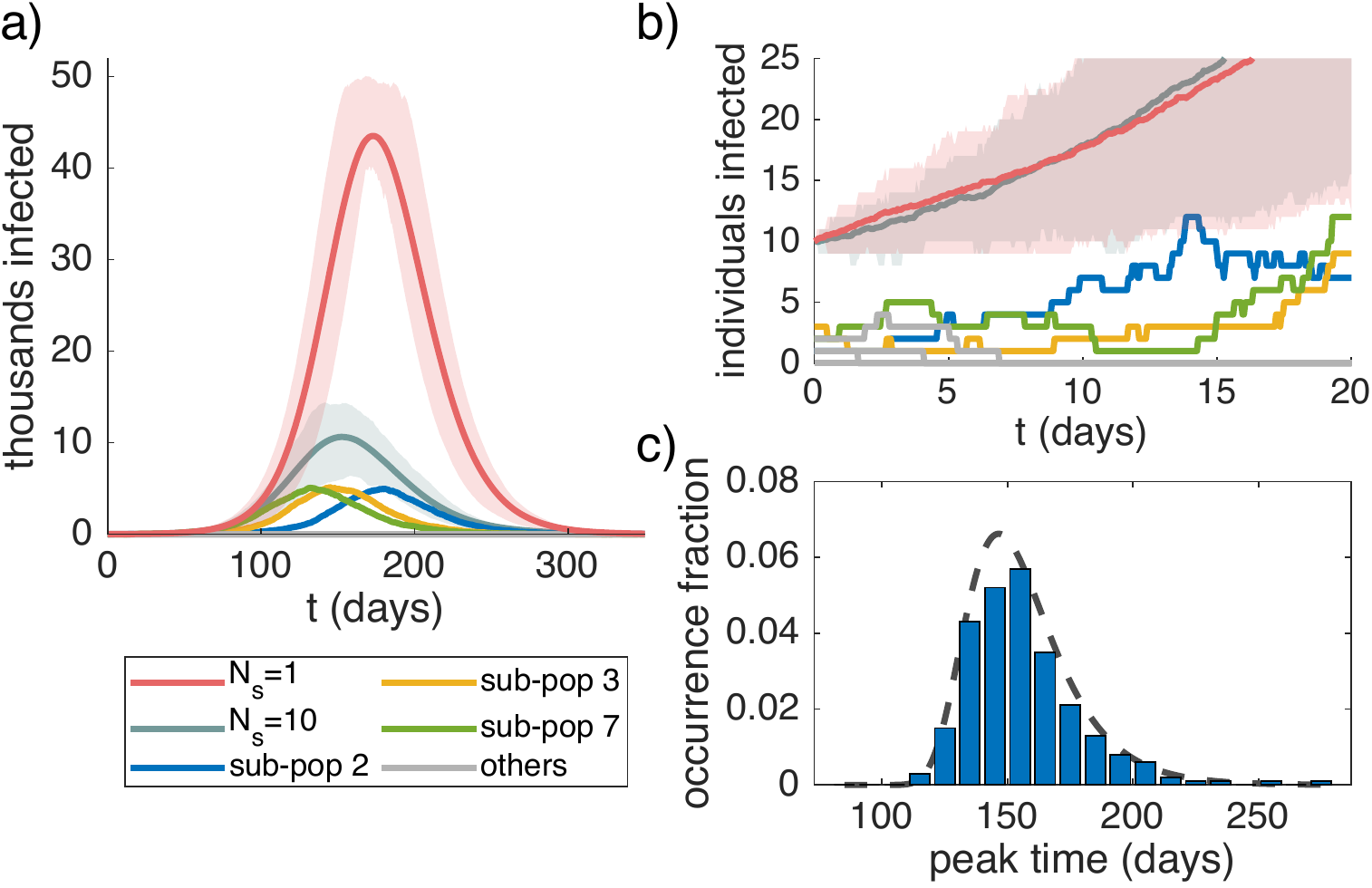}
\caption{{Stochastic effects lower the peak in subdivided populations.} (a)~Time course for a population of $N=\ $1,000,000 with $I_0=10$ initially infected individuals for $N_s=1$ large population (red) and a population split into $N_s=10$ sub-populations (turquoise), $b=0.2$, $k=0.14$. Shading indicates $\pm 25\%$ confidence intervals across 100 simulations. Sub-populations are shown from one simulation with $N_s=10$. (b)~Enlarged plot of the initial phase for the same traces as in panel a. (c)~Distribution of peak times in sub-populations for $N_s=10$. Occurrence fraction indicates fraction of sub-populations across all simulations. Dashed line indicates analytical approximation, Eq.~\eqref{eq:peakdistribana}, with uniform $n=I_0/N_s=1$.}
\label{fig:examplesubpop}
\end{figure}

\section{Mathematical model}
\label{sec:model}
\subsection{Reaction system}
\label{sec:reactionsystem}
We consider a population of $N$ individuals with SIR dynamics~\cite{Kermack:1927cy}
\begin{equation}
\label{eq:stochasticModel}
    {\rm S}+{\rm I}\xrightarrow{\  b/N\ } {\rm I}+{\rm I}, \hskip 1cm {\rm I}\xrightarrow{\  k\ }{\rm R}, %\emptyset
\end{equation}
with S, I and R referring to {\em susceptible}, {\em infected} and {\em removed} individuals, respectively, where removal with per-capita rate $k$ happens due to recovery, quarantine or death. The rate $b$ corresponds to the number of contacts per unit time an individual has with a random other individual in the population, multiplied by the probability that a contact between a susceptible and an infected individual leads to transmission. The total transition rate from S to I per unit time is therefore $\frac{b}{N}\,S\,I$. The two rates $b$ and $k$ are related to the basic reproduction number $\cR_0=b/k$, which is independent of population size. The deterministic epidemic threshold above which an outbreak occurs is $\cR_0=1$, and we assume $\cR_0>1$ throughout this study. The population is subject to the total constraint $N=S(t)+I(t)+R(t)$, where we denote the number of individuals in each state by the same letters. For simplicity, all initial conditions assume $R(0)=0$, such that they are uniquely defined by $N$ and the number of initially infected $I_0 = I(0)$.

When a population of total size $N$ is split up into $N_s$ sub-populations, we simulate $N_s$ separate copies of the system~\eqref{eq:stochasticModel}, with $N$, $S$, $I$ and $R$ replaced by $N_i$, $S_i$, $I_i$ and $R_i$ respectively, where the index $i$ refers to the different sub-populations and $N_i=N/N_s$. $N=\sum N_i$, $S=\sum S_i$, $I=\sum I_i$ and $R=\sum R_i$ refer to the population totals. The initial number of infected individuals are distributed either uniformly or randomly across the $N_s$ sub-populations. All numerical results in this study are obtained from stochastic simulations of Eq.~\eqref{eq:stochasticModel} using the Gillespie algorithm~\cite{Gillespie:1977dc}. To account for the inherent stochasticity of the system, several realizations, i.e. identical simulations with different random number generator seeds, are simulated for each parameter set. We report the number of realizations as well as distributions, averages and standard deviations across the results as appropriate. Our main figure of interest is the peak number of infected individuals $I_{\max}$ or equivalently the peak infected fraction of the population $\gamma=I_{\max}/N$. These could be considered a measure for the impact of the epidemic and the strain on the health care system and public health resources such as the agencies that perform contact tracing and testing.

\subsection{Deterministic behavior}
The reaction scheme \eqref{eq:stochasticModel} results in the deterministic mean-field equations
\begin{subequations}
\label{eq:deterministic}
\begin{align}
    \frac{\textrm{d} S}{\textrm{d} t} &= -\frac{b}{N}\,S\,I,   \\
    \frac{\textrm{d} I}{\textrm{d} t} &= \frac{b}{N}\,S\,I-k\,I,\\
    \frac{\textrm{d} R}{\textrm{d} t} &= k\,I,
\end{align}
\end{subequations}
which give rise to two regimes in the dynamics. During the initial regime, $I$ starts off from an initial value $I(0)=I_0$, rises exponentially $\sim I_0 e^{(b-k)t}$, and saturates to a peak value 
\begin{equation}
I_{\max} \equiv \gamma N\approx \left(1-\frac{k}{b}\big[1+\log(b/k)\big]\right)N, \label{eq:peakdet}
\end{equation}
where the approximation for the maximum fraction of infected individuals $0<\gamma<1$ is valid as long as the entire population is initially susceptible, i.e. $S(0)\approx N$~\cite{weissSIR}. In the secondary regime when the recovery dynamics dominates, $I$ decays to zero exponentially, as the number of susceptibles decreases below the value necessary to sustain spreading. 

In this deterministic system, a subdivision into $N_s$ smaller sub-populations of size $N/N_s$ will have no effect, since Eq.~\eqref{eq:deterministic} remains invariant when $S$, $I$, $R$ and $N$ are scaled by the common factor $1/N_s$. Relative to their individual sub-population sizes $N_i$, the same dynamics are therefore observed in all sub-populations and the dynamics of the population totals $S=\sum S_i$, $I=\sum I_i$ and $R=\sum R_i$ are identical to those of a single large population. The subdivision is therefore not analogous to cutting links in a contact network but rather a redistribution of them, since we assume that the contact rate $b$ remains unchanged. This conservative assumption means that individuals in each sub-population still have the same number of contacts per unit time as they had in the large population, despite the smaller number of individuals to choose from. While, in reality, the contact rate $b$ might decrease in such a situation and deterministically reduce $\cR_0$ and therefore $I_{\rm max}$, we intentionally keep it constant here to extract the effects of stochasticity.

\subsection{Stochastic behavior}
Deterministic behavior only applies if $S$ and $I$ are both large, particularly only after the number of infected people $I$ has risen to appreciable levels. If $I$ is still low, stochastic fluctuations determine whether $I$ will ``take off'' and develop exponential behavior, even if $b>k$. This effect was already considered shortly after Kermack and McKendrick introduced the original SIR model~\cite{Whittle:1955kn} and is now well-known. However, in a subdivided population, it can significantly alter the course of the outbreak in the total population if the initial number of infected individuals in a single sub-population is low enough (even if the number is large in the total population). An example for populations of $N=\ $1,000,000 individuals split into $N_s=10$ sub-populations is shown in Figs.~\ref{fig:examplesubpop}a and \ref{fig:examplesubpop}b, along with the expected dynamics of a single large population (red curve). In one example set of 10 sub-populations, only three sub-populations (blue, yellow, and green curves) experience a significant outbreak, and they are desynchronized with a broad distribution of individual peak times (\ref{fig:examplesubpop}c). Spontaneous extinction and desynchronization lead to an average behavior across 100 simulations with a significantly reduced peak (turquoise curve). Note that, on average, both the undivided large population and the sum of the smaller sub-populations initially exhibit comparable exponential growth in the number of infected individuals (Fig.~\ref{fig:examplesubpop}b). This means that, while extinction in some sub-populations and fluctuations in timing happen early on, their effect is only seen later during the saturation phase.

During the initial phase, we can assume that $S\approx N$ and that $I$ follows a simple birth-death process with rates $b$ for ``birth'' and $k$ for ``death''. We shall use this analogy for derivations throughout this study and in the appendices. We briefly recapitulate one important result from the theory of branching processes here, namely that an exponentially growing population that starts from an initial condition of $I(0)=1$ has a finite {\it extinction} probability of 
\beq
\cP_0 (t)=\frac{k}{b} \cdot \frac{e^{(b-k) t}-1}{e^{(b-k) t}-k/b} \quad .\label{eq:P0}
\eeq
which asymptotically approaches $k/b$ at long times; see the derivation in Appendix~\ref{sec:birth-death}. This means that with probability $p_{1}^{\rm ext}=k/b$ the dynamics never enters the exponentially growing deterministic regime, but decays back to zero due to number fluctuations\footnote{Note that, for simplicity of presentation, we are using the long-time limit of the extinction probability to define $p_{1}^{\rm ext}$ rather than $\cP_0(t)$. This simplification implies that extinction happens fast enough on the time scale relevant for our problem. It can be justified by comparing the time scale of the extinction process to that of the infection peak in the SIR model (see further below in the main text and Appendix).}. For two independent lineages in the same population, the extinction probability is therefore $p_{2}^{\rm ext}=(k/b)^2$, and, similarly, $p_{n}^{\rm ext}=(k/b)^n$, as long as the total population is sufficiently large such that the lineages do not interfere with each other. We will use these extinction probabilities and other statistics of the birth-death process to derive analytical approximations for the effects of extinction and desynchronization on the stochastic dynamics.

\section{Results}
\subsection{Theoretical estimates for isolated sub-populations}
\subsubsection{Extinction}
To obtain an estimate for the effect of extinction and the distribution of infected individuals, we add up the maximum numbers of infected individuals in the sub-populations. Each of these peaks is approximately $\gamma N/N_s$, but only if the infection does not stochastically become extinct during the initial stages. For large population sizes and values of $b/k$ that result in a significant peak, extinction usually happens well before the peak is reached in other sub-populations (see Appendix~\ref{sec:timescale}), such that these populations do not contribute.  Therefore, on average, the contribution of each sub-population will be $I_{s,{\rm max}}(n) = \gamma\left(1-p_{n}^{\rm ext}\right) {N}/{N_s}$,
where $n$ indicates the number of initial infected individuals in the sub-population and $p_{n}^{\rm ext}$ is the probability that they go extinct without entering deterministic growth as discussed above. Therefore, the total peak number of infected individuals in all the sub-populations due to extinction is given by $I_{{\rm max}}^{\rm ext} = \sum_{n} g_n I_{s,{\rm max}}(n)$, 
where $g_n$ is the number of sub-populations with $n$ initially infected individuals. Note that $N_s = \sum_{n}g_n$.
Combining the above equations, we obtain
\begin{align}\label{eq:Iextmaxgeneral} 
I_{{\rm max}}^{\rm ext} = \gamma\,\frac{N}{N_s}\sum_{n} \left(1-p_{n}^{\rm ext}\right) g_n=\gamma N \left[1-\frac{\sum_{n} g_n (k/b)^n}{\sum_{n} g_n }\right].
\end{align}
The above result manifestly shows that 
\begin{align}\label{eq:gammaext}
\gamma^{\rm ext} \equiv \frac{I_{{\rm max}}^{\rm ext}}{N} =\gamma \left[1-\sum_{n} \frac{g_n}{N_s} (k/b)^n\right]< \gamma
\end{align}
holds. Note that this reduction is exclusively due to extinction and the simple summation of the individual maxima neglects the possible desynchronization between sub-populations, which we will consider further below.

\begin{figure}[t]
\centering
\includegraphics[width=\columnwidth]{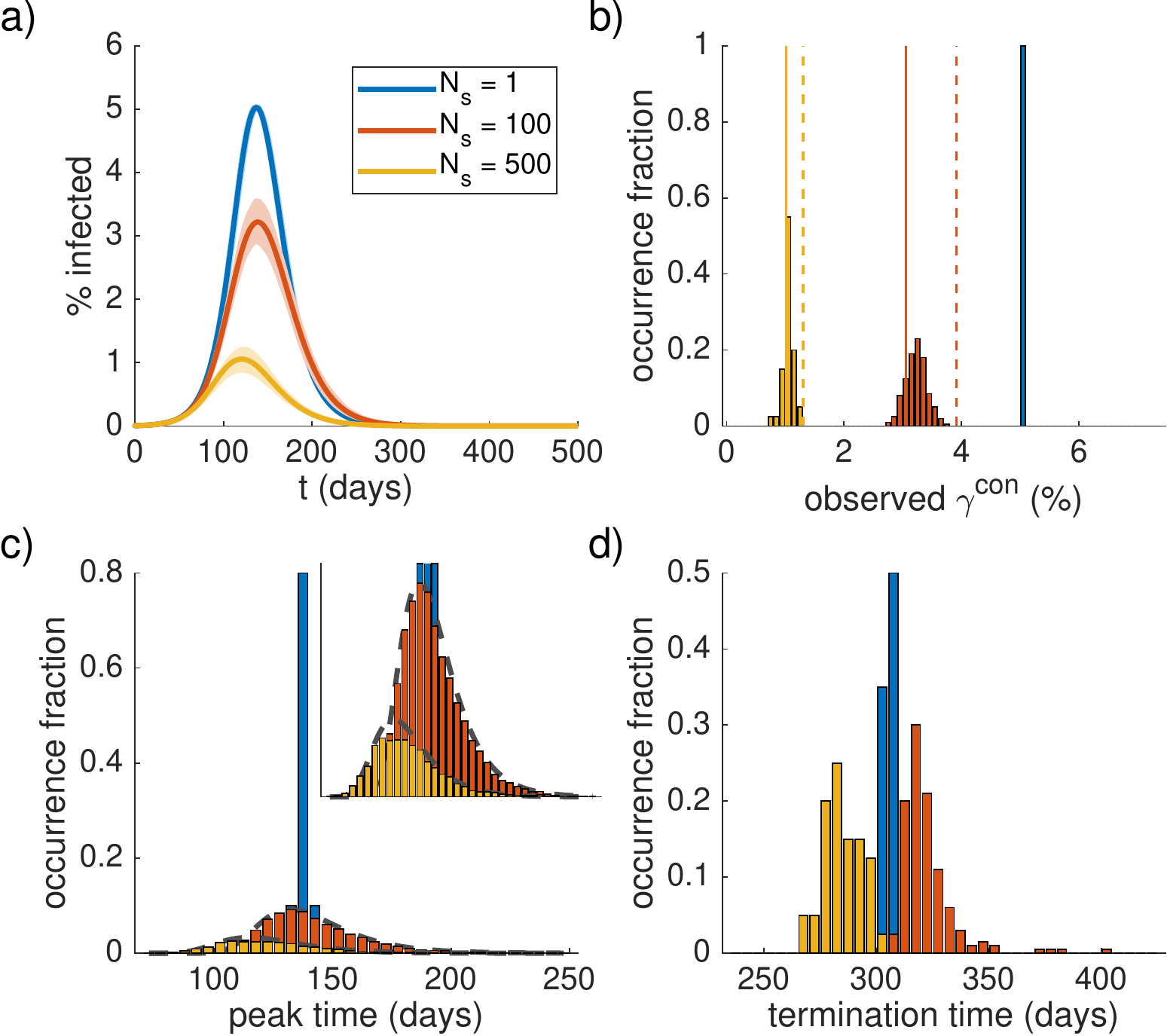}
\caption{{Epidemics for different subdivisions of the population.} $N=\ $8,000,000, $b=0.2$, $k=0.14$, three different values of $N_s$. 20, 200 and 40 individual simulations for $N_s=1$, $N_s=100$ and $N_s=500$, respectively. (a) Time courses (solid lines) and 2.5/97.5 percentiles (shading). (b) Distributions of observed peak percentage $\gamma^{\rm con}$ (in the whole population). Occurrence fraction indicates fraction of simulations. Analytical estimates for $\gamma^{\rm ext}$ (dashed lines), Eq.~\eqref{eq:gammaext}, and $\gamma^{\rm con}$ (solid lines), Eq.~\eqref{eq:gammacon}, assume $g_n$ according to a binomial distribution. (c)~Distribution of peak times in the sub-populations. Inset provides an enlarged y-axis. Dashed lines indicate analytical approximation, Eq.~\eqref{eq:peakdistribana}, assuming a uniform $n=I_0/N_s$ for each case. (d)~Distribution of termination times, defined as the time when $I$ in the total population drops below $I_0$.}
\label{fig:LS}
\end{figure}

For example, for the ideal case where each sub-population only contains at most one infected individual, we have
\begin{align}
 \gamma_1^{\rm ext} = \gamma\, \frac{I_0}{N_s} \left(1-\frac{k}{b}\right),
\end{align}
where $g_1=I_{0}$ is the total number of initially infected individuals in the large population (for this to make sense, $N_s\geq I_0$ is required). Since $\gamma$ corresponds to the case where the population was not split up, the peak number of infected can therefore be reduced by increasing the number of sub-populations $N_s$ or by bringing $b$ closer to $k$. Note that this is \emph{in addition} to a potential decrease in the deterministic peak fraction $\gamma$ of infected (cf. Eq.~\eqref{eq:peakdet}) that would result if the subidivision also led to fewer contacts (i.e. a reduced rate $b$), which we have conservatively assumed not to be the case here.

\subsubsection{Desynchronization}
The independent summation of maxima in different sub-populations is a conservative estimate, since fluctuations can lead to stochastic desynchronization and thus to a further reduction of the peak value. The distribution of peak times in the sub-populations from the previous example is shown in Fig.~\ref{fig:examplesubpop}c. The temporal shift between the different sub-populations can be attributed entirely to stochastic fluctuations in the initial phase of the dynamics. Assuming that this time shift accumulates while the dynamics can still be modeled as a pure birth-death process without saturation effects, we can derive the probability distribution for the deviation from the mean peak time $\Delta t_{\rm peak}\equiv t_{\rm peak}-\aver{t_{\rm peak}}$ as
\beqa
\label{eq:peakdistribana}
&&P(\Delta t_{\rm peak})=k (1-{k}/{b})[1-(k/b)^n] \nonumber \\
&&\times\exp\left(-(b-k)\left(\bar\tau+\Delta t_{\rm peak}\right)-\frac{k}{b} \,e^{-(b-k)\left(\bar\tau+\Delta t_{\rm peak}\right)}\right),\nonumber\\
\eeqa
where $n$ is the initial number of infected individuals in the sub-population and $\bar\tau=\ln\left(\gamma^\prime k/b\right)/(b-k)$ with $\gamma^\prime$ being the exponential of the Euler constant (see Appendix~\ref{sec:analyticalpeakdistrib} for details). Note that $n$ here was only used to incorporate the extinction probability, while the shape of the distribution is based on a single initially infected individual. Nevertheless this result is in excellent agreement with the measured distribution for randomly distributed infected individuals (see dashed line in Fig.~\ref{fig:examplesubpop}c).

\begin{figure}[t]
\centering
\includegraphics[width=\columnwidth]{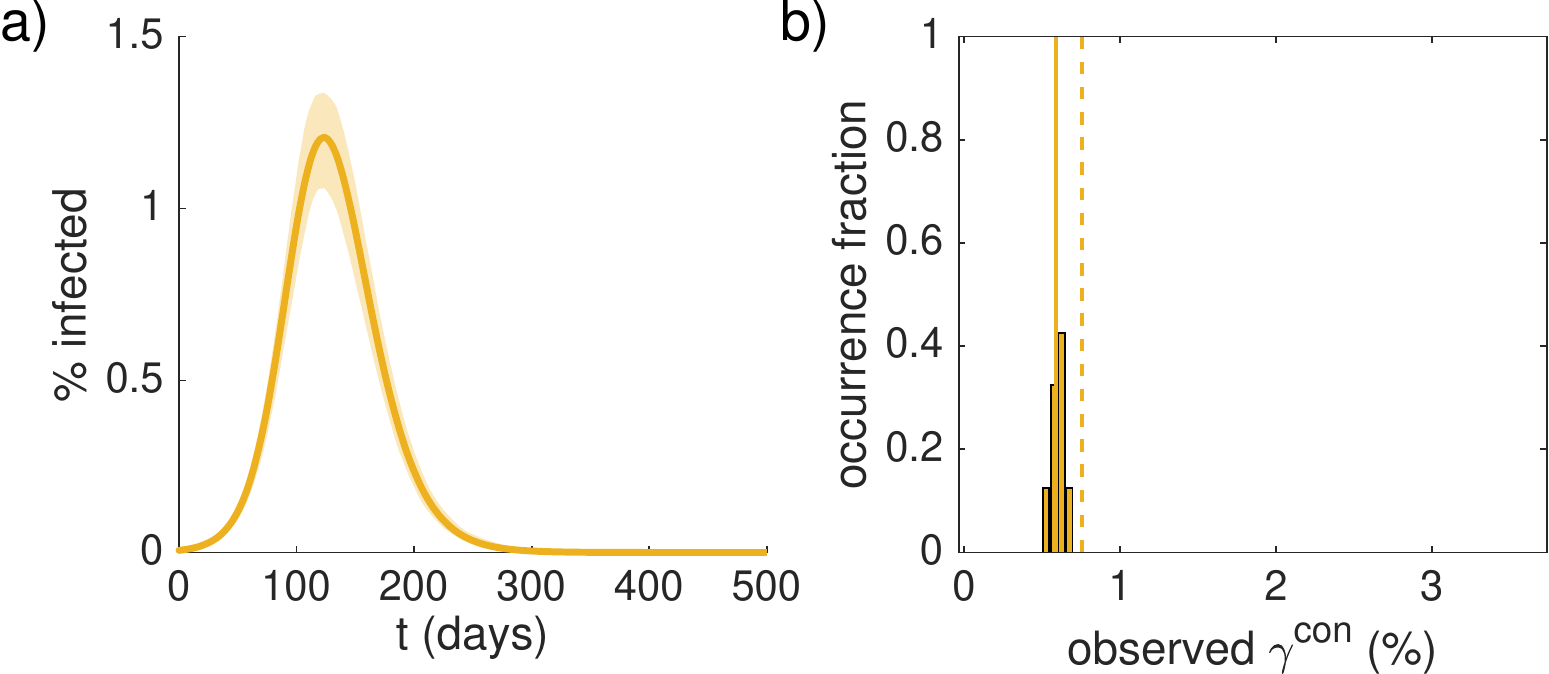}
\caption{Plots analogous to Fig.~\ref{fig:LS}a and b for the case $N_s=500$, but with exactly one initially infected individual in each sub-population instead of a random distribution. Analytical estimates, Eq.~\eqref{eq:gammaext} (dashed line) and Eq.~\eqref{eq:gammacon} (solid line), accordingly use $g_1=I_0=500$.}
\label{fig:LSsplit500exactly1perpop}
\end{figure}

We can then use this distribution to obtain a quantitative estimate for the additional peak reduction due to desynchronization. For this purpose, we approximate the deterministic time evolution of $I$ in the vicinity of the peak as
$I(t)\approx N\gamma\exp\left(-\frac{1}{2}\,bk\gamma\,(t-t_\text{peak})^2\right)$,
which is valid as long as $S(t)$ remains of order $\sim N$ (see Appendix~\ref{sec:desynceffect}), i.e. $b/k$ is not too large. In the limit of many superimposed peaks of this shape, with the variability of $t_\text{peak}$ given by Eq.~\eqref{eq:peakdistribana}, the peak is reduced by an additional factor $\alpha^{-1}$:
\begin{equation}
    \label{eq:gammacon}
\gamma^{\rm con}=\frac{\gamma^{\rm ext}}{\alpha},\quad\alpha=\sqrt{1+\frac{\pi^2[\cR_0-1-\log(\cR_0)]}{6(\cR_0-1)^2}}.
\end{equation}
The peak number of infected individuals, with both stochastic effects of the {\it confinement} taken into account, similarly becomes $I^{\rm con}_{\rm max}=N\gamma^{\rm con}={I^{\rm ext}_{\rm max}}/{\alpha}$.
It is interesting to note that this reduction factor is bounded from below by $\lim_{\cR_0\rightarrow 1}\alpha^{-1} = \sqrt{12/(12+\pi^2)}\approx0.7407$. The desynchronization effect is therefore much more limited than the extinction effect.

\begin{figure}[t]
\centering
\includegraphics[width=\columnwidth]{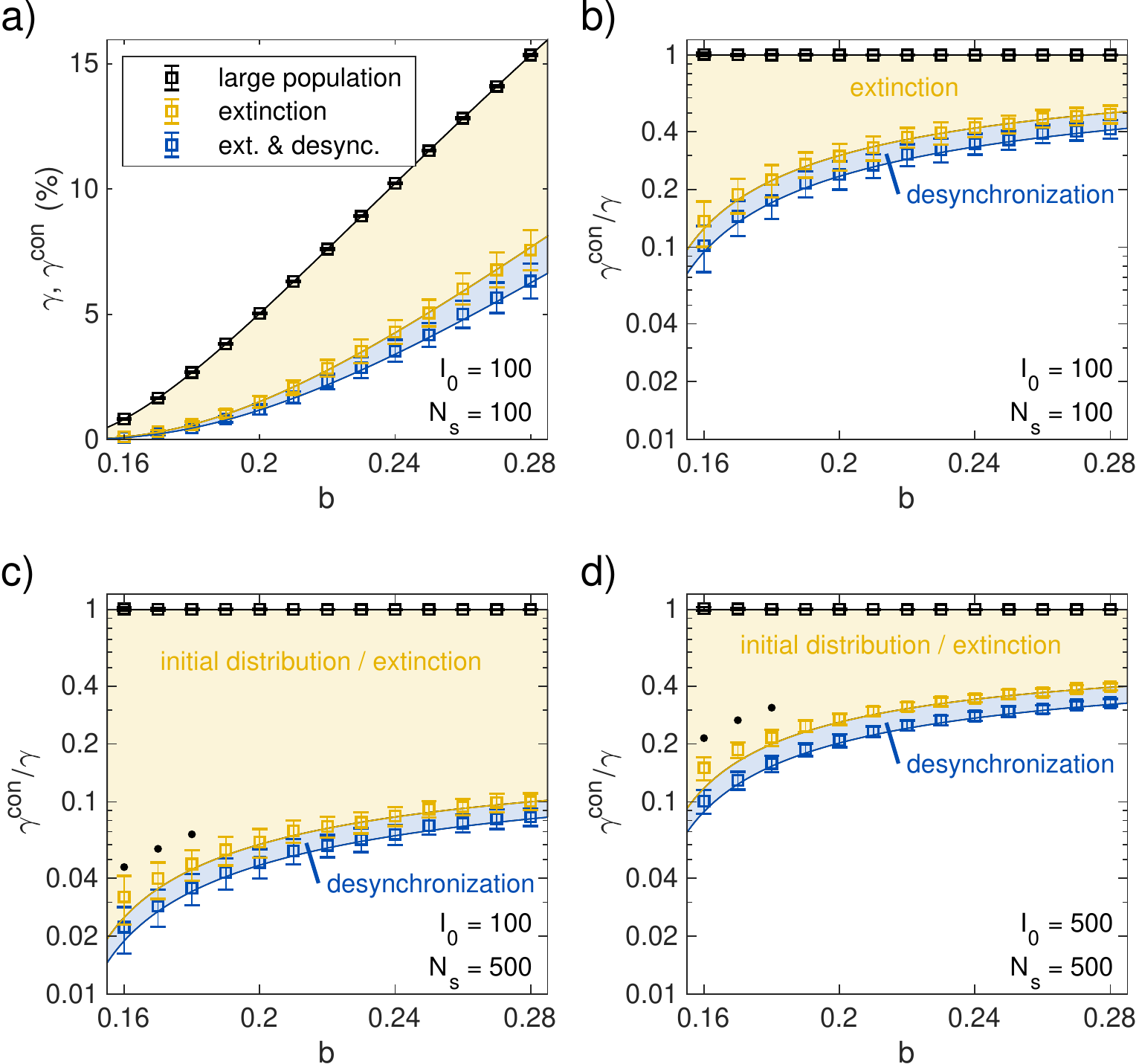}
\caption{{Peak reduction for different subdivisions and values of $b$.} $N=\ $8,000,000, $k=0.14$. All data points represent avarages across 100 independent stochastic simulations, errors bars indicate standard deviation. (a)~Peak fraction of infected individuals for  $I_0=100$, $N_s=100$ with each sub-population initially containing exactly one infected individual. Symbol colour indicates reduction due to extinction (yellow) or both extinction and desynchronization (blue) as measured in simulations. Black symbols represent the large-population control. Yellow/blue shading and solid lines indicate analytical predictions from Eq.~\eqref{eq:gammaext} and Eq.~\eqref{eq:gammacon}, respectively. Black line indicates deterministic estimate from Eq.~\eqref{eq:peakdet}. (b) Same as panel (a), plotted logarithmically and normalized by theoretical peak fraction $\gamma$ without subdivision, Eq.~\eqref{eq:peakdet}. (c) $I_0=100$, $N_s=500$, each sub-population containing at most one infected individual. Yellow color now represents the reduction both to extinction and the initial distribution of infected individuals, since 400 sub-populations are initialized with $I=0$. (d) $I_0=500$, $N_s=500$, infected individuals randomly distributed across sub-populations. Black dots in (c) and (d) mark data points where the estimation of the extinction effect is affected by overlapping timescales between different processes (see text and Appendix~\ref{sec:timescale}).}
\label{fig:maximaVSb}
\end{figure}

\subsection{Numerical results}
We consider as an example a region with a population of 8,000,000 
and 500 infected individuals ($I_0/N \sim 6\cdot10^{-5}$),
and assume a removal rate of $k=0.14$, corresponding to a realistic mean removal time of $1/k\approx7$~days for the recent epidemic~\cite{Ferretti:2020iq} (particularly if symptomatic individuals are quickly removed from the infectious pool through quarantining). Let us further assume that the infectious contact rate is $b=0.2$ ($> k$). This corresponds to a substantial reduction of $\cR_0$ from its initial value of 2--2.5~\cite{Li:2020ct} through mild measures such as social distancing, although the epidemic would still spread exponentially, with infection numbers doubling about every 12 days.

If this population is allowed to mix homogeneously, the dynamics will evolve according the deterministic prediction with a peak around $5\%$ infected individuals (blue data in Figs.~\ref{fig:LS}). If instead, the population is split up and the 500 infected people are distributed randomly across the sub-populations, the peak percentage of infected individuals decreases to around $3\%$ (for 100 sub-populations of 80,000 people) or $1\%$ (for 500 sub-populations of 16,000 people) on average (red and yellow, respectively). In all cases, the analytical estimate which only considers the extinction effect, Eq.~\eqref{eq:gammaext}, is only an upper bound for the peak percentage of infected individuals in the total population, while also considering desynchronization according to Eq.~\eqref{eq:gammacon} yields a good estimate the typical peak values. The peak time distributions for the three different ways of splitting up the population shown in Fig.~\ref{fig:LS}c also agree with the analytical estimate of Eq.~\eqref{eq:peakdistribana}. Note that these distributions are not normalized since a significant fraction of sub-populations experience extinction of the epidemic and therefore do not exhibit a peak. There is also a subtle, non-monotonic effect on the termination time of the epidemic (Fig.~\ref{fig:LS}d), whose distribution is broader when the population is split up, but does not change position appreciably. Note that the reduction for $N_s=500$ sub-populations in Fig.~\ref{fig:LS} is comparable (or even slightly lower) than the case where the 500 infected individuals are not distributed randomly across the sub-populations, but each sub-population contains exactly one infected individual. In this case (see Fig.~\ref{fig:LSsplit500exactly1perpop}), 
there are no sub-populations with initially zero infected individuals, implying that the reduction in peak value compared to the large homogeneous population is strictly due to extinction and desynchronization, which are again well predicted by the analytical estimates.

To examine the validity of our approximations across different parameters, we varied the contact rate $b$ and carried out numerical simulations for values of $\cR_0$ ranging between 1.14 and 2. We analyzed the resulting peak magnitudes to extract the individual contributions of extinction and desynchronization, which are in excellent agreement with our predictions of Eq.~\eqref{eq:gammaext} and \eqref{eq:gammacon}, as shown in Fig.~\ref{fig:maximaVSb}. The contribution of extinction alone was estimated numerically by summing maxima in different sub-populations, regardless of their timing. Overall, the simulations confirm the relative importance of the extinction effect, whereas the additional reduction by desynchronization plays a smaller role. Figs.~\ref{fig:maximaVSb}a and b show the case where $N_s=I_0=100$, i.e. number of sub-populations and initially infected individuals is the same, and exactly one infected individual is placed in each sub-population. This serves to demonstrate the maximum effect of extinction, whereas in Fig.~\ref{fig:maximaVSb}c, a large share of the peak reduction is due to sub-populations containing no infections, as $I_0=100<N_s=500$. However, random distribution of infected individuals for $N_s=I_0=500$ (Fig.~\ref{fig:maximaVSb}d) leads to a very similar result as in Fig.~\ref{fig:maximaVSb}b, although some of the reduction is due to the initial distribution (i.e. sub-population without any infections). For a high number of sub-populations $N_s$ as in Figs~\ref{fig:maximaVSb}c and d (and consequently smaller sub-population size), deviations from the theory begin to appear towards low values of $b$ very close to $k$, as the time scale of the extinction process becomes comparable to that of the deterministic SIR dynamics. In this regime, the distinction between an initial stochastic phase approximated by a birth-death process and the onset of saturation effects becomes increasingly blurred, as we show analytically in Appendix~\ref{sec:timescale}. In particular, this affects the estimation of the extinction contribution (marked by black dots).

\section{Discussion}
Reducing the infectious contact rate $b$ or increasing the removal rate $k$ directly leads to a decrease of the deterministic peak fraction of infected, $\gamma$.
The above analysis shows that, even without changing $\cR_0=b/k$, the isolation of small sub-populations can reduce the overall peak number of infected people in the ideal case of at most one infected individual per sub-population by an additional factor of up to $I_0/N_s\cdot(1-k/b)/\alpha$ when $I_0/N_s<1$. One contribution comes from the communities which have no infections and are now protected ($I_0/N_s$), while another contribution comes from the possibility that an infection chain in the local community stochastically ends due to fluctuations ($k/b$). Stochastic desynchronization ($1/\alpha$) further reduces the peak by up to about 25\% according to Eq.~\eqref{eq:gammacon}. However, as shown by our estimates and confirmed by the numerical simulations, even outside this ideal scenario, a reduction can be achieved, regardless of the distribution of infected individuals across the sub-populations, and the reduction will be larger if $b/k$ is already close to 1. It is also worth noting that, in contrast to reductions in $\cR_0=b/k$, the time scale of the outbreak is not increased. 

The benefits of subdivision are obvious even from a deterministic standpoint in the case where many regions initially contain no infected individuals --- in this case, subdivision prevents spreading of the epidemic to disease-free communities. However, our analysis shows that this advantage persists due to stochastic extinction events and desynchronization even if the sub-populations are so large that many or all of them initially contain infections, as long as $I_0/N_s\sim 1$. Of course, increasing $N_s$ further is always beneficial due to the above-mentioned deterministic effect, with the trivial limiting case of one group per household (an extremely strict lockdown). In contrast, aiming at $I_0/N_s\sim 1$ could still allow for a functioning local socioeconomic life in fairly large sub-populations, if $I_0$ is not too large when the subdivision happens.

While extinction has been widely considered for SIR-type models~\cite{Whittle:1955kn,Andersson:2000wk} and has been related to a minimum number of infections necessary to cause a ``major'' outbreak~\cite{Allen:2017jm}, we have shown here that, even if the dynamics in the large population is outside the stochastic regime, it is possible to resurrect these effects by artificially sub-dividing the population.
Because of the strong exponential dependence of the extinction probability on $n$ (see Eq.~\eqref{eq:Iextmaxgeneral}) it is important to note that $I_0$ denotes the \emph{true} number of infections, including undetected and/or asymptomatic cases. Another aspect we have neglected here is that of cross-infections: In reality, sub-populations cannot be perfectly isolated, and so local extinction might only be temporary, as has been seen in studies of persistence~\cite{Keeling:2000gf,Andreasen:1989hf}. The calculated peak reduction would be observed in the limit of small cross-infection rates. In contrast to extinction, desynchronization does not reveal itself on the level of a single population (except as a difference in timing) and is therefore an emergent property of the subdivision scenario, which is likely to persist in the presence of cross-infections. In the framework presented in section~\ref{sec:reactionsystem}, these could be included (without changing $\cR_0$) by allowing a certain fraction $\xi$ of contacts across the entire population, and only restricting the remaining fraction $1-\xi$ to \emph{within} each sub-population. We set up such a model in a separate study~\cite{Bittihn:2020do} to investigate a potential realistic containment strategy.

In reality, individuals will not compensate for all avoided contacts outside the local sub-population with contacts within it, as we have a conservatively assumed by keeping $b$ constant upon subdivision. Instead, isolation will naturally lead to a reduction in $b$, akin to cutting links in the spreading network~\cite{Matamalas:2018ck}, so that the effect of subdivision will be a combination of deterministic reductions in $\cR_0$ and the stochastic effects presented here. Subdivision of a population can be complementary to containment measures, such as social distancing and electronic contact tracing~\cite{Ferretti:2020iq,Huerta:2002fe}, which still allow for a functioning local public life. However, it also does not preclude the activation of more drastic measures in regions beginning to show deterministic exponential behavior~\cite{Bittihn:2020do}.

\section*{Data availability statement}
All information to generate the data that supports the findings of this study is available within the article.

\begin{acknowledgments}
We have benefited from discussions with J. Agudo-Canalejo, A. Bahrami, H. Bickeboeller, E. Bodenschatz, W. Brück, H. Chat{\'e}, R. Fleischmenan, T. Friede, T. Geisel, H. Grubmüller, R. Jahn, B. Mahault, V. Priesemann, T. Richter, S. Scheithauer, A. Vilfan, M. Wilczek, and R. Yahyapour. The research was supported by the Max-Planck-Gesellschaft.
\end{acknowledgments}

%\bibliography{bibliography}


%\bibliographystyle{plain}
% \begin{thebibliography}{}

% \end{thebibliography}

\begin{appendix}

\section{Exact solution of the birth-death process}
\label{sec:birth-death}

Consider a population of the infected individuals $I$ that can undergo the following two processes:
\beq
{\rm I} \; \xrightarrow{b} \; {\rm I}+{\rm I}, \quad \quad \quad {\rm I} \; \xrightarrow{k} \; \emptyset,\label{eq:process-AA0}
\eeq
i.e., each I can give birth to another I with rate $b$, or, it can die with rate $k$, at any time. Ignoring the stochasticity, the average behavior of the system is described by exponential birth and death. The population $\bar{n}(t)$ can be determined as follows:
\beq
\dt{\bar{n}(t)}=(b-k) \bar{n}(t) \quad \Rar \quad \bar{n}(t)=e^{(b-k) t},
\eeq
where we have assumed that the initial size of the population is one. As this is a one-step process, the probability of finding $n$ copies of $I$ in the sample at time $t$ satisfies the following Master equation
\beq
\dt{\cP_n (t)}  = k (n+1)\; \cP_{n+1}(t) + b (n-1) \; \cP_{n-1}(t) - \left(k + b \right) n \, \cP_n(t),\label{eq:birth-death-Meq}
\eeq
The factor of $n$ is needed because the birth or death could happen to anyone. Equation (\ref{eq:birth-death-Meq}) can be solved by an {\it ansatz} of the form $\cP_n \sim f^n$ for $n \geq 1$, which together with the initial condition $\cP_n (0)=\delta_{n,1}$ gives us the solution as
\beq
\cP_n (t)=\frac{\bar{n} (1-k/b)^2}{(\bar{n}-1)(\bar{n}-k/b)} \; \left(\frac{\bar{n}-1}{\bar{n}-k/b}\right)^n. \label{eq:Pn-birth-death}
\eeq

The distribution can be used to calculate the first two moments
\beqa
\aver{n(t)} &=& \sum_{n}^\infty n \, \cP_n (t)=\bar{n}(t)=e^{(b-k) t}, \\
\Delta n^2 &=& \aver{\left[n-\aver{n}\right]^2}=\left(\frac{b+k}{b-k}\right) \; e^{(b-k) t} \left[e^{(b-k) t}-1\right],\nonumber \\
\eeqa
which reveal more interesting features about the system. First, it is reassuring that the average population size behaves according to the mean-field description above that predicted exponential growth or decay. A quantity of interest is
\beq
\frac{\Delta n^2}{\bar{n}}=\left(\frac{b+k}{b-k}\right) \left[e^{(b-k) t}-1\right],
\eeq
which probes whether number fluctuations follow a characteristic Poisson behavior. In the long time limit, we have
\beq
\frac{\Delta n^2}{\bar{n}}=\left\{ \begin{array}{ll}
      \displaystyle  \infty \quad & \mbox{; \quad$b > k$},\\ \\
\displaystyle
        \frac{k+b}{k-b} & \mbox{; \quad $b < k$},
\end{array} \right.
 \label{eq:fluctuations}
\eeq
which shows that while a decaying population that corresponds to $b < k$ has a Poisson behavior, a growing population corresponding to $b> k$ has {\em giant number fluctuations}, which can be characterized via
\beq
\frac{\Delta n}{\bar{n}}= \sqrt{\frac{b+k}{b-k}} \; \sqrt{1-e^{-(b-k) t}},
\eeq
which leads to
\beq
\frac{\Delta n}{\bar{n}}= \sqrt{\frac{b+k}{b-k}},
\eeq
in the long time limit. In other words, the fluctuations scale with the average population size when $b> k$, and with the square root of the average population size when $b < k$.

The above solution allows us to calculate {\it the extinction probability} of the population $\cP_0 (t)$, which is an {\it absorbing state}. We find
\beq
\cP_0 (t)=1-\sum_{n=1}^\infty \cP_n (t)=\frac{k}{b} \cdot \frac{e^{(b-k) t}-1}{e^{(b-k) t}-k/b}.\label{eq:P0-birth-death}
\eeq
which is a very interesting result. When $k > b$, $\bar{n} \to 0$ at long times, and we obtain $\cP_0=1$. It is no surprise that extinction at long times is a certainty when the death rate is larger than the birth rate. However, when $k < b$, $\bar{n} \to \infty$ at long times, and we obtain $\cP_0=k/b$; a result that is in contradiction with the prediction of the average behavior of the system, which is exponential growth. So, number fluctuations could completely annihilate an exponentially growing population.

\section{Time scale of the extinction process and accuracy of maxima detection in sub-populations}
\label{sec:timescale}
Here, we derive quantitative estimates that allow us to compare the time scale of the extinction process to that of the deterministic peak in the SIR model. This is conceptually interesting in its own right, but it also allows us to meaningfully differentiate between ``real'' maxima and random transient peaks in the number of infected individuals in sub-populations that experience extinction.

In the pure birth-death process, the fraction of extinction events, $0\leq\phi_{\rm x}\leq1$, that have already happened by time $t$ can easily be calculated from Eq.~\eqref{eq:P0-birth-death} as
\beq
\phi_{\rm x}(t)=\frac{\cP_0(t)}{\lim_{t\rightarrow\infty}\cP_0(t)}=1-\frac{1-k/b}{e^{(b-k)t}-k/b}.
\eeq
This equation can be inverted to yield the time $t_{\rm x}$ by which a fraction $\phi_{\rm x}$ of extinction events have happened:
\beq
t_{\rm x}(\phi_{\rm x})=\frac{1}{b-k}\log\left({\frac{1-\phi_{\rm x} k/b}{1-\phi_{\rm x}}}\right).
\eeq
On the other hand, we can also estimate the fraction of non-extinct populations, $0\leq\phi_c\leq 1$, that will still be below a cutoff size $n_c$ at time $t$:
\begin{align}
\nonumber\phi_c(t, n_c) &= \sum_{n=1}^{n_c-1}{\cP_n(t)},\\
&=1-\frac{e^{(b-k)t}-k/b}{e^{(b-k)t}-1}\left(1-\frac{1-k/b}{e^{(b-k)t}-k/b}\right)^{n_c}.
\end{align}
Evaluating $\phi_c(t_{\rm x}(\phi_{\rm x}),n_c)$ therefore yields the fraction of populations still below $n_c$ when a fraction $\phi_{\rm x}$ of extinction events have already happened. This expression can be inverted to yield the simple relationship
\beq
n_c(\phi_{\rm x},\phi_c)=1+\frac{\log(1-\phi_c)}{\log \phi_{\rm x}},
\eeq
giving the number of infected individuals below which a fraction $\phi_c$ of non-extinct populations will still be, at the time when a fraction $\phi_{\rm x}$ of populations destined for extinction have already reached the extinct state.

In order to estimate the effect of extinction in our numerical simulations (cf. Fig.~\ref{fig:maximaVSb}), we detect the maximum number of infected individuals in each sub-population (independent of their timing), and compare the sum of these numbers to our estimate $I_\text{max}^\text{ext}$ from the main text. In the sub-populations that experience random extinction of the epidemic, the detected numerical maxima will in reality be transient fluctuations before extinctions. These contribute more and more as $\cR_0=b/k\rightarrow 1$, when the deterministic peak value $N\gamma=N \big(1-(1+\log{\cR_0})/\cR_0 \big)$ \cite{weissSIR} decreases and the extinction probability $1/\cR_0$ increases. Using the estimates above, we can exclude these false maxima based on their timing, by only considering those maxima for which
\beq
t_{\rm max}>t_{\rm x}(\phi_{\rm x}),
\eeq
and simultaneously ensuring that
\beq
n_c(\phi_{\rm x},\phi_c)<N\gamma,\label{eq:peaknotreached}
\eeq
is fulfilled. $\phi_{\rm x}$ and $\phi_c$ play the role of accuracy parameters. The first condition ensures that false maxima are excluded with probability $\phi_{\rm x}$, while the second one ensures that a pure birth-death process would not have reached the deterministic SIR peak by the same time with probability $\phi_c$. Note that the latter is a conservative estimate, as growth in the SIR model is significantly slowed before reaching its peak compared to a pure birth-death process. In Fig.~\ref{fig:maximaVSb}, we use a value of $\phi_{\rm x}=\phi_c=0.99$ to exclude 99\% of false maxima and still detect more than 99\% of deterministic SIR maxima, except for the data points marked as unreliable, for which  Eq.~\eqref{eq:peaknotreached} is not fulfilled and therefore the extinction process and the deterministic SIR peak are not clearly separated in time. Conversely, this also means that for all other parameters (i.e. larger $\cR_0=b/k$), extinction usually happens well before the deterministic SIR dynamics reaches its peak.

It is worth emphasizing that, in the limit $b\rightarrow k$ and small populations, the distinction between an initial stochastic phase and a deterministic time course becomes meaningless, since $\gamma$ eventually becomes order $\sim1/N$ and the mean extinction time diverges. At this point, the dynamics throughout will be dominated by random growth of the number of infected individuals and stochastic fluctuations will continue to contribute, even as the number of susceptibles decreases, eventually ending the epidemic (i.e. during and beyond the maximum). In addition, the assumption that there is no depletion of susceptibles in the early phase (and thus the equivalence to a pure birth-death process) breaks down. However, in this study, we are interested in the regime where even sub-populations are still large, and while $b$ is sufficiently close to $k$ to yield a significant extinction probability $k/b$, it is large enough to lead to a significant deterministic outbreak peak. Therefore, we do not investigate this regime.

\section{Analytical approximation of the relative peak time distribution}
\label{sec:analyticalpeakdistrib}
The fact that the early phase of the dynamics in the SIR model (when $S\approx N$ and $I$ is small) corresponds to a simple birth-death process also allows us to obtain an analytical estimate for the peak time distributions of the sub-populations. This can be readily adapted from a similar calculation performed on an equivalent problem in evolution, where the dynamics of a small mutant sub-population with a given selective advantage can likewise be understood as a birth-death branching process \cite{Desai:2007fh}, for which the transition from the initial stochastic regime where extinction is still possible to the deterministic regime of exponential growth corresponds to the establishment of the mutation in the population (which precedes fixation). 

We obtain an approximation for the establishment time distribution of the disease in a sub-population as
\beq
\label{eq:P-SIR}
P^{\rm est}_{\rm SIR}(\tau)=k \left(1-{k}/{b}\right)\exp\left(-(b-k)\tau-\frac{k}{b} \,e^{-(b-k)\tau}\right),
\eeq
%|\,{\rm not\ extinct}
where we have corrected for an additional minus sign missing from Ref.~\cite{Desai:2007fh}. The variation in the timing of the later deterministic dynamics is due entirely to fluctuations in this initial stochastic phase.
To compare this analytical approximation with our simulation results for the peak time in the main text, we plot the non-normalized, unconditional distribution
\beq
\label{eq:peaktimedistrib}P(t_{\rm peak}) =[1-(k/b)^n] P^{\rm est}_{\rm SIR}\left(t_{\rm peak}+\bar \tau-\langle t_{\rm peak}\rangle\right),
\eeq
which is diminished by a factor $[1-(k/b)^n]$ (from Eq. \eqref{eq:P-SIR}) accounting for the probability of extinction in a population with initially $n$ infected individuals, and has its mean shifted to the measured mean peak time $\langle t_{\rm peak}\rangle$. Here
\beq
\bar\tau\equiv\langle\tau\rangle=\frac{1}{b-k}\,\ln\left(\gamma^\prime \;\frac{k}{b}\right),
\eeq
where $\gamma^\prime=1.7810724\cdots$ is the exponential of Euler's constant. 

We note that simply shifting the mean of the distribution is justified because the dynamics is predominantly identical in different sub-populations once they are in the deterministic regime, while only lagging by a random time span $\tau$. This simple argument depends on the assumption that stochastic fluctuations can be ignored before deviations from exponential behavior (i.e. saturation effects) have to be considered for the deterministic dynamics. This is true for the scenarios we consider in the SIR model, since our sub-populations still consist of thousands of individuals and we are explicitly focusing on cases where $b$ is not arbitrarily close to $k$.\\

\section{Estimating the effect of sub-population desynchronization}
\label{sec:desynceffect}
For estimating the peak reduction effect due to desynchronization of the sub-populations, it is convenient to work with the normalized equations for $s=S/N$ and $i=I/N$, which read
\begin{subequations}
\begin{align}
    \dot s &= -bsi, \label{eq:snorm}\\
    \dot i &= bsi -ki \label{eq:inorm}.
\end{align}
\end{subequations}
When $i$ reaches its peak $\gamma=i(t_\text{peak})$, new infections and recovery balance according to Eq. \eqref{eq:inorm} and $s(t_\text{peak})=k/b$. Based on this known value, we use the following {\it ansatz} for $s$
\beq
s(t) = \frac{k}{b}\big(1+\varepsilon(t)\big),
\eeq
with $\varepsilon(t_\text{peak})=0$. Since we are interested in the regime where there is a substantial extinction probability $k/b$, $s(t_\text{peak})$ is also still of order 1. Together with the fact that $\dot i(t_\text{peak})=0$ by definition, we expect from Eq.~\eqref{eq:snorm} that the lowest (linear) order of $\varepsilon$ will suffice to describe the dynamics around the peak, i.e. $\varepsilon(t)\approx\varepsilon_1\cdot(t-t_\text{peak})$ (conversely, we expect this approximation to break down when $b\gg k$). Substituting the {\it ansatz} into Eq.~\eqref{eq:snorm} yields $\varepsilon_1=-b\gamma$, or
\beq
\varepsilon(t)\approx -b\gamma(t-t_\text{peak}).
\eeq
With this, we can obtain an approximation for $i$ around the peak. From Eq.~\eqref{eq:inorm}, we know that
\beq
\dt{\log(i)} = bs(t)-k=k\varepsilon(t),
\eeq
which can easily be solved. Together with the condition $i(t_\text{peak})=\gamma$, we obtain
\beq
i(t)\approx\gamma\exp\left(-\frac{1}{2}bk\gamma(t-t_\text{peak})^2\right).\label{eq:inearpeak}
\eeq
Now that we have an approximation for $i(t)$ near the peak, we can calculate how these time courses add up across individual sub-populations, by assuming that they all have the shape \eqref{eq:inearpeak}, with the peak time $t_\text{peak}$ stochastically distributed according to Eq.~\eqref{eq:peaktimedistrib}. Defining
\beq
\bar i(t)= \lim_{N_s\rightarrow\infty}\frac{1}{N_s}\sum_{j=1}^{N_s}i(t^j_\text{peak};t),
\eeq
where each $i(t^j_\text{peak}; t)$ represents a time course as in Eq.~\eqref{eq:inearpeak} with $t^j_\text{peak}$ drawn from the distribution \eqref{eq:peaktimedistrib} for each $j$, we obtain an average superposition of many sub-populations in the limit $N_s\rightarrow\infty$:
\beq
\label{eq:superposintegral}\bar i(t)=\int \mathrm{d}t_\text{peak}\,P^\text{est}_\text{SIR}(t_\text{peak}+\bar\tau)\, i(t_\text{peak}; t).
\eeq
Note that (as compared to Eq.~\eqref{eq:peaktimedistrib}) we use here the normalized distribution, without the diminishing factor due to extinction, in order to extract the reduction strictly due to desynchronization. We have also set $\langle t_\text{peak}\rangle$ to 0 without loss of generality, as a different value would simply shift $\bar i(t)$ by the corresponding time.

The integral in Eq.~\eqref{eq:superposintegral} cannot be integrated in closed form. We therefore replace $P^\text{est}_\text{SIR}$ by a normal distribution $\mathcal{N}(0,\sigma^2)$ with the same variance $\sigma^2=\pi^2/[6(b-k)^2]$. It is useful to note that, as for the normal distribution, the variance completely determines the shape of the Gumbel distribution in Eq.~\eqref{eq:P-SIR}, which means that the systematic error introduced by this replacement is parameter independent. Finally, we can calculate
\begin{align}
\nonumber\bar i(t)&=\int \mathrm{d}t_\text{peak}\,\frac{\exp\left({-{t_\text{peak}^2}/(2{\sigma^2})}\right)}{\sqrt{2\pi}\sigma}\, i(t_\text{peak}; t), \\
\label{eq:superposresult}&=\frac{\gamma}{\alpha}\exp\left(-\frac{1}{2}bk\gamma \frac{t^2}{\alpha^2}\right),\\
\nonumber&\quad\text{with }\alpha=\sqrt{1+\frac{\pi^2 bk\gamma}{6(b-k)^2}}.
\end{align}
The maximum of the resulting time course occurs at $t=0$ (due to our arbitrary choice of the mean for $t_\text{peak}$) and is $\bar i(0)=\gamma/\alpha$. Since the expected peak value without desynchronization is $\gamma$, desynchronization reduces this peak value by a factor of $\alpha^{-1}$. According to Eq.~\eqref{eq:superposresult}, $\alpha$ itself depends on $\gamma$, which in turn is a function of $\cR_0=b/k$. Using the well known approximation $\gamma=1-[1+\log(\cR_0)]/\cR_0$ \cite{weissSIR}, which is valid as long as $S\approx N$ initially, we rewrite $\alpha$ as
\beq
\alpha=\sqrt{1+\frac{\pi^2[\cR_0-1-\log(\cR_0)]}{6(\cR_0-1)^2}}.
\eeq
While we expect the quantitative estimate to be less accurate towards higher $\cR_0$ (see above), we note that the important limits
\begin{align}
\lim_{\cR_0\rightarrow\infty}\frac{1}{\alpha} &= 1,\\
\lim_{\cR_0\rightarrow 1}\frac{1}{\alpha} &= \sqrt{\frac{12}{12+\pi^2}}\approx0.7407,
\end{align}
exist. The first one signifies that there is no peak reduction due to desynchronization for $\cR_0\rightarrow\infty$, consistent with the disappearance of the stochastic phase at the beginning of the dynamics. The second limit indicates a finite reduction by a factor $\approx0.7407$ towards $\cR_0=b/k=1$. Since the time scales of both the stochastic fluctuations and the deterministic peak behavior diverge for $\cR_0\rightarrow1$ (and are ill-defined for $\cR_0=1$), this means that they must exhibit identical scaling behavior in order for neither of them to dominate. In between the two extremes, $1/\alpha$ increases monotonically with $\cR_0$, which implies that the maximum reduction that can be achieved by desynchronization is about 26\% and is reached close to $\cR_0=1$. It is important to note, however, that several assumptions even about the deterministic time course (for example, the value of $\gamma$) break down when $\cR_0$ is so close to 1 that $\gamma$ becomes of order $1/N$, so a fully stochastic treatment would be needed to fully capture this regime. This does not limit the validity of the results in the regime we are interested in, i.e. where sub-populations still exhibit clear deterministic outbreaks (or extinction).\\

\end{appendix}

\end{document}